\documentclass{article}
\usepackage{spconf}

\usepackage{graphicx}
\usepackage{subcaption}
\usepackage[dvipsnames]{xcolor}
\usepackage{multirow}
\usepackage{booktabs,colortbl}
\usepackage{arydshln}
\usepackage{scalerel}
\usepackage[utf8]{inputenc} % allow utf-8 input
\usepackage[T1]{fontenc}    % use 8-bit T1 fonts
\usepackage[hidelinks]{hyperref}       % hyperlinks
\usepackage{url}            % simple URL typesetting
\usepackage{booktabs}       % professional-quality tables
\usepackage{amsfonts}       % blackboard math symbols
\usepackage{nicefrac}       % compact symbols for 1/2, etc.
\usepackage{microtype}      % microtypography
\usepackage{wrapfig}
\usepackage{mathtools}
\usepackage{romannum}

\usepackage{listings}
\usepackage{geometry}
\definecolor{nopass}{HTML}{333333}
\definecolor{mycolor}{HTML}{d4d42a}
\colorlet{aliased}{mycolor!90!black}
\definecolor{aliasedtangled}{HTML}{ff0000}
\definecolor{nonaliased}{HTML}{2a2ad4}

\title{\vspace{-5pt}How convolutional neural networks deal with aliasing\vspace{-1pt}}
\name{Antônio H. Ribeiro$^{\dagger,\star}$, Thomas B. Sch\"on$^{\dagger}$\thanks{This work was financially supported by: the Brazilian research agency CAPES, the \emph{Wallenberg AI, Autonomous Systems and Software Program (WASP)} funded by Knut and Alice Wallenberg Foundation, the Swedish Foundation for Strategic Research (SSF) via the project \emph{ASSEMBLE} (contract number: RIT15-0012) and by the \emph{Kjell och M{\"a}rta Beijer Foundation}.}\vspace{-5pt}}
\address{
$^{\dagger}$  Department of Information Technology, Uppsala University, Sweden\\
$^{\star}$ Department of Computer Science, Universidade Federal de Minas Gerais, Brazil
\vspace{-3pt}}

\begin{document}

% Resilience

\maketitle

\begin{abstract}
The convolutional neural network (CNN) remains an essential tool in solving computer vision problems. Standard convolutional architectures consist of stacked layers of operations that progressively downscale the image.  Aliasing is a well-known side-effect of downsampling that may take place: it causes high-frequency components of the original signal to become indistinguishable from its low-frequency components. While downsampling takes place in the max-pooling layers or in the strided-convolutions in these models, there is no explicit mechanism that prevents aliasing from taking place in these layers. Due to the impressive performance of these models, it is natural to suspect that they, somehow, implicitly deal with this distortion. The question we aim to answer in this paper is simply: ``how and to what extent do CNNs counteract aliasing?'' We explore the question by means of two examples: In the first, we assess the CNNs capability of distinguishing oscillations at the input, showing that the redundancies in the intermediate channels play an important role in succeeding at the task; In the second, we show that an image classifier CNN while, in principle, capable of implementing anti-aliasing filters, does not prevent aliasing from taking place in the intermediate layers.

\end{abstract}


\section{Introduction}

In signal processing applications, low-pass filters are often placed before the downsampling operations to prevent \textit{aliasing} from happening. These filters remove the high-frequency components that would otherwise be distorted during the downsampling operation and become indistinguishable from the low-frequency signal components.

While, earlier CNN models do implement rudimentary low-pass filters before subsampling by means of \textit{average pooling}~\cite{le_cun_handwritten_1990}. Average pooling has gradually lost space to the \textit{max-pooling layer}, which shows a better task performance~\cite{scherer_evaluation_2010}, and to \textit{strided-convolutions}, which became very popular and that are used in architectures such as the residual neural network~\cite{he_deep_2016}. However, neither of these mechanisms provide an explicit way of preventing aliasing from occurring. 

\begin{figure}
 \vspace{-4pt}
 \centering
    \begin{tabular}{cccc}
    \hspace{-15pt}
    \begin{subfigure}{0.12\textwidth}
        \centering
        \smallskip
        \includegraphics[height=1\linewidth]{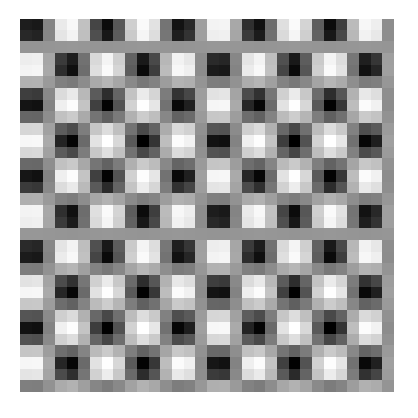}
        \caption{$\left(\frac{\pi}{3}, \frac{\pi}{3}\right)$}
    \end{subfigure}
    &
    \hspace{-15pt}
    \begin{subfigure}{0.12\textwidth}
        \centering
        \smallskip
        \includegraphics[height=1\linewidth]{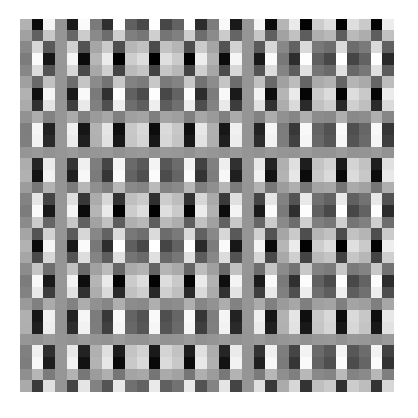}
        \caption{$\left(\frac{2\pi}{3}, \frac{\pi}{3}\right)$}
    \end{subfigure}
    &
    \hspace{-15pt}
    \begin{subfigure}{0.12\textwidth}
        \centering
        \smallskip
        \includegraphics[height=1\linewidth]{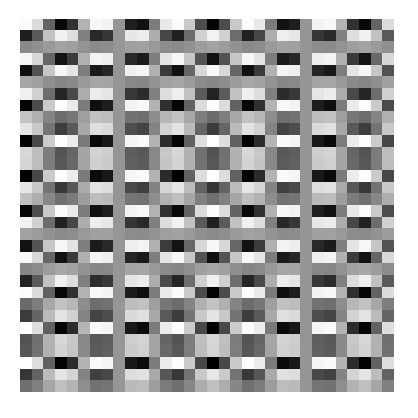}
        \caption{$\left(\frac{\pi}{3}, \frac{2\pi}{3}\right)$}
    \end{subfigure}
    &
    \hspace{-15pt}
    \begin{subfigure}{0.12\textwidth}
        \centering
        \smallskip
        \includegraphics[height=1\linewidth]{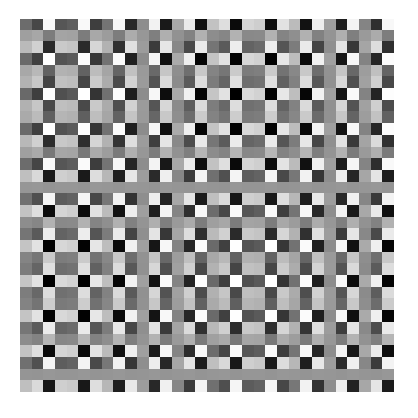}
        \caption{$\left(\frac{2\pi}{3}, \frac{2\pi}{3}\right)$}
    \end{subfigure}
    \end{tabular}
    \vspace{-5pt}
    \caption{\textbf{Four oscillatory patterns}. If the image is downsampled by a factor of two, the patterns become \textit{indistinguishable due to aliasing}. The patterns are generated by Eq.~(\ref{eq:oscilatory_2d}), with $A = 0.01$. The oscillation frequencies $(\omega_1, \omega_2)$ are given in the sub-captions.}
    \label{fig:oscilatory_2d}
    \vspace{-14pt}
\end{figure}

The rather natural idea of introducing anti-aliasing filters before the downsampling operations in CNNs has been explored recently \cite{zhang_making_2019, azulay_why_2019, tanaka_wavecyclegan2_2019, henaff_geodesics_2016, vasconcelos_effective_2020, zou_delving_2020}. For instance, in the context of image classification, Zhang \cite{zhang_making_2019} presented a successful implementation of a CNN with anti-aliasing filters, showing that such an approach yields CNNs invariant to shifts, with extra robustness and, even, improved accuracy. 

%% PREPRINT (comment out when it is not the case)
\begin{figure}[b] 
\vspace{-5pt}
\noindent\rule[0.5ex]{\linewidth}{1pt}
\noindent \footnotesize{ Preprint. To appear in the 2021 IEEE International Conference on Acoustics, Speech and Signal Processing (ICASSP).  Please cite:}
\begin{lstlisting}
@inproceedings{ribeiro_how_2021,
author={Ant\^onio H. Ribeiro and Thomas B. Sch\"on},
title={How Convolutional Neural Networks Deal with Aliasing}, 
year={2021},
publisher={IEEE}, 
booktitle={2021 {IEEE} International Conference on Acoustics, Speech and Signal Processing, {ICASSP}}
}
\end{lstlisting}
\vspace{-5pt}
\end{figure}

Our work, however, takes a different route: rather than trying to explicitly include anti-aliasing mechanisms in a CNN, we aim to explain how these architectures still manage to be successful without it and what mechanisms are implicitly learned to counteract aliasing. 

Aliasing results in loss of information and the inability to distinguish between frequency components. The first question we address is: "does this happen in a CNN?" This question is studied by means of a toy example, designed to indicate whether a convolutional neural network can distinguish between oscillatory components at its input, and what role do the network width and depth play in it. 

In the second experiment, we go one step further and try to assess if CNNs actually prevent aliasing from taking place in the intermediate layers. Convolutional layers are, in principle, capable of implementing anti-aliasing filters. Hence, even if these are not hard-coded in the structure, CNNs could learn it. The experiment tries to establish whether this happens for image classification problems.

\vspace{3pt}
\begin{small}
\noindent
\textbf{Code Availability:} The code for reproducing the experiments is available at: \href{https://github.com/antonior92/aliasing-in-cnns}{https://github.com/antonior92/aliasing-in-cnns}.
\end{small}
\vspace{-5pt}

\section{Toy example: classifying oscillations}
\label{sec:classifying-oscilations}

In this section, we describe a simple toy example that requires the model to learn to distinguish between oscillatory components. In CNNs, the components are downsampled in the intermediate layers and hence might suffer aliasing. We try to establish whether this behavior in the intermediate layers affects the global ability of the CNN to distinguish between frequencies at its input.

\subsection{Task description}
The model receive as input a $32$ by $32$ oscillatory two-dimensional signal $x$ and need to classify it according to the frequency $\omega = (\omega_1, \omega_2)$. Here, each component $\omega_i$ is the frequency of the oscillations in one direction. The training and validation sets consist of pairs $\{x, \omega\}$, where $\omega$ belongs to the set ${\{(\frac{k\pi}{N} , \frac{l\pi}{N} )\mid k, l \in \{0, \cdots, N-1\}\}}$, which contain $N^2$ uniformly spaced frequencies in ${[0, \pi)\times[0, \pi)}$. And, the input $x$ is oscillatory, with its $(i,j)$-th entry
\begin{equation}
    \label{eq:oscilatory_2d}
    x_{i, j} = \cos(\omega_1 i + \theta_1)\cos(\omega_2 j + \theta_2) + A \alpha_{i, j},
\end{equation}
where the phases $\theta_1$ are $\theta_2$ are sampled from uniform distributions: $\theta_1 \sim \mathcal{U}(0, 2\pi)$, $\theta_2\sim \mathcal{U}(0, 2\pi)$. The signal is  also corrupted by additive white noise, $\alpha_{i, j}\sim \mathcal{U}(-1/2, 1/2)$, generated independently for each $i$ and $j$. The amplitude of this additive white noise is $A$. Fig.~\ref{fig:oscilatory_2d} show samples generated by this procedure.

Every time the signal $x$ is downsampled by a factor of two, the number of frequencies that can be distinguished drops by a factor of four due to aliasing. For instance, the oscillatory patterns in Fig.~\ref{fig:oscilatory_2d} would become indistinguishable after being downsampled by a factor of 2. For the CNN models, however, these downsampling operations happen at intermediate layers and the goal of this task is to assess what is the resolution CNNs can resolve between distinct frequency components at the input.

\subsection{Results}

\begin{figure}
\vspace{-10pt}
\includegraphics[width=1.05\columnwidth]{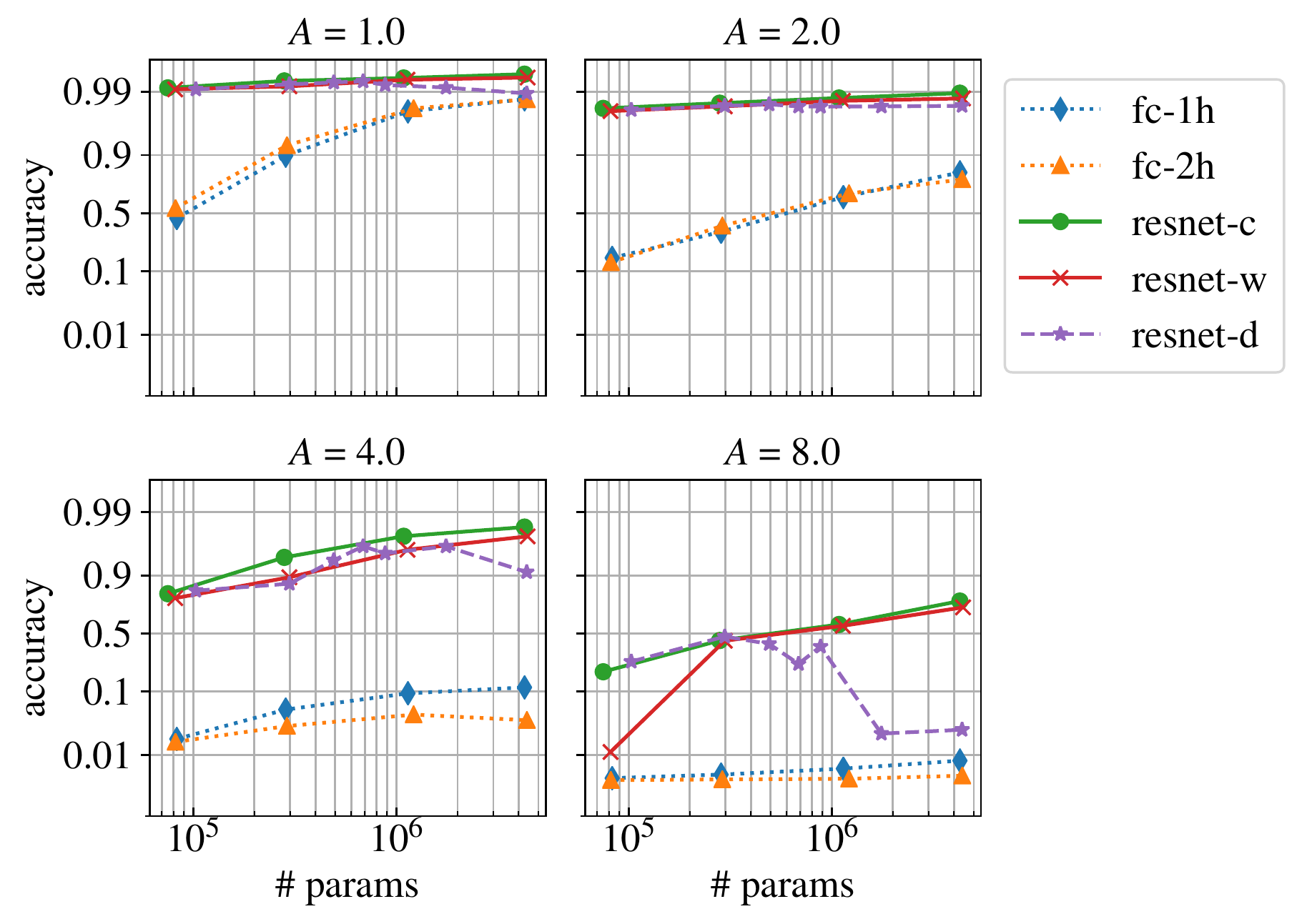}
\caption{\textbf{Accuracy \textit{vs} number of parameters.} Plot showing accuracy for models of different sizes. \texttt{fc-1h} and \texttt{fc-2h} are fully connected neural networks with one and two hidden layers, respectively. For the ResNet we increase the number of parameters by fixing the depth and changing the number of channels in the intermediate layers: 1) keeping the number of channels constant across the convolutional layers (\texttt{resnet-c}); or 2) with the number of channels doubling at every block where downsampling takes place  (\texttt{resnet-w}). We also compare it with: 3) keeping the channels constant and increasing the depth (\texttt{resnet-d}). The $y$-axis is in logit scale and the $x$-axis in log scale to facilitate the visualization.}
\label{fig:acc_vs_size}
\vspace{-6pt}
\end{figure}

In Fig.~\ref{fig:acc_vs_size}, we  compare the accuracy of the residual neural network (ResNet)~\cite{he_deep_2016} with that of fully connected neural networks.  The reported accuracy is on the test set of $10\,000$ samples and was obtained for models trained on $20\,000$ samples. The samples were generated, independently, using Eq.~(\ref{eq:oscilatory_2d}). We set $N=20$ for generating a uniform grid of frequencies with $N^2 = 400$ possibilities. Both the training and test sets are balanced, with each one of the $400$ classes occurring the same number of times.  The models have been trained during $50$ epochs using a batch size of $128$, the Adam optimizer~\cite{kingma_adam:_2014} with default betas, a weight decay of $10^{-4}$ and an initial learning rate of $0.001$, which is decreased by a factor of ten in epoch 35. 

The ResNet implementations have two downsampling operations taking place in the forward direction. Hence, without any way of circumventing aliasing, which might occur during downsampling, the CNNs would perform poorly at the task. The fact that they perform better than the fully connected networks (which do not downsample the image) strongly suggests that CNNs can indeed somehow circumvent aliasing. Furthermore, as the problem gets harder due to extra noise, increasing the number of channels can improve the performance. Increasing the depth, however, does not help and can even degrade performance. We hypothesize that, while the signals in individual channels might suffer from aliasing, the redundancy allows the information to still be recovered by the neural network in later stages. This mechanism is further investigated in the following sections.

This task has the same spirit as the experiments in~\cite{gong_impact_2018}. There they perform, in a uni-dimensional signal setup, experiments assessing the ability of CNNs to distinguish between different input frequencies. However, while in their experiments, the CNN model fails to effectively distinguish between frequency components, ours succeeds at it. We hypothesize that one reason is the low number of filters in the intermediate layers used in their model. These filters are critical for succeeding in the task (i.e., Fig.~\ref{fig:acc_vs_size}). Their choice of model architecture, which stacks temporal and 2-D convolutional layers, might also help to explain the differences in the results.

\section{Background}
\label{sec:background}

\subsection{The discrete Fourier transform}
\label{sec:dft}

There is a one-to-one correspondence between the signal and the amplitudes and phases of the sine components it can be decomposed into. For discrete signals with finite length such a representation is obtained by means of the \textit{Discrete Fourier Transform} (DFT). Let $X$ denote the DFT of an image, i.e. a \textit{two-dimensional} signal $x \in\mathbb{R}^{m \times n}$. We represent this transformation by: $x \xleftrightarrow{\text{DFT}} X,$ where $X\in \mathbb{C}^{m \times n}$ is a matrix of complex numbers. Rewriting this matrix on polar form results in $X = A e^{i \Phi}$, where $A$ and $\Phi$ are matrices in $\mathbb{R}^{m \times n}$  representing the amplitude and the phase of the complex entries in $X$. 

The amplitude and the phase of the DFT have an easy interpretation: the $(k, l)$-th entry of $A$ is the amplitude and the corresponding entry of $\Phi$ is the phase shift of the oscillatory component at the frequency $\omega = \left(\frac{2\pi}{n} k, \frac{2\pi}{m} l\right)$. Here, the frequency $\omega$ is represented by a tuple, with each element of the tuple corresponding to the frequency in one dimension of the two-dimensional signal.

\subsection{Downsampling and aliasing}
\label{sec:downsampling-and-aliasing}

\begin{figure}[t]
    \vspace{-7pt}
    \hspace{10pt}
    \begin{subfigure}{0.47\textwidth}
        \centering
        \includegraphics[width=0.8\linewidth]{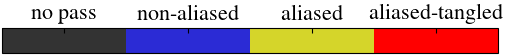}
        \vspace{-1pt}
    \end{subfigure}
    
    \begin{tabular}{cc}
    \hspace{10pt}
        \begin{subfigure}{0.21\textwidth}
            \centering
            \includegraphics[width=1\linewidth]{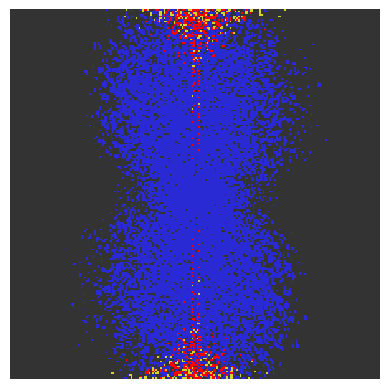}
            \caption{\textcolor{gray}{65\%}, \textcolor{blue}{33\%}, \textcolor{ForestGreen}{1\%}, \textcolor{red}{1\%}}
        \end{subfigure}
        \hspace{-10pt}
        &
        \begin{subfigure}{0.21\textwidth}
            \centering
            \includegraphics[width=1\linewidth]{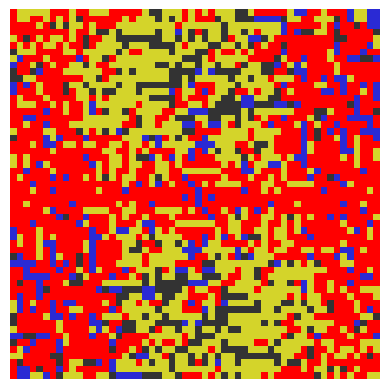}
            \caption{\textcolor{gray}{15\%}, \textcolor{blue}{10\%}, \textcolor{ForestGreen}{35\%}, \textcolor{red}{39\%}}
        \end{subfigure}
    \end{tabular}
    \vspace{-5pt}
    \caption{\textbf{Aliasing in CNNs}. DFT points classified according to Section~\ref{sec:aliasing_methodology} for the intermediate signals of the ResNet34 evaluated on an ImageNet test sample. The signals correspond to intermediate tensors immediately after the strided convolutions. The DFT has been rearranged in such a way that low frequencies appear at the center of the image.}
    \label{fig:aliasing_in_CNNs}
    \vspace{-10pt}
  \end{figure}

For simplicity, let $r$ be an integer and assume $m$ and $n$ to both be multiples of $r$. Given $x\in \mathbb{R}^{m \times n}$, let $x'\in \mathbb{R}^{\frac{m}{r} \times \frac{n}{r}}$ denote the signal $x$ downsampled by a factor $r$. The $(k, l)$-th entry of $x'$ is given by
\begin{equation}
    \label{eq:downsampling}
    x'_{k, l} = x_{r\,k, r\, l}.
\end{equation}
In turn, the DFT $X'$ of $x'$ is related to that of $x$,
\begin{equation}
    \label{eq:downsampling_fft}
    X' = \frac{1}{r^2} \sum_{i=1}^r \sum_{j=1}^r X^{(i, j)},
\end{equation}
where $X^{(i, j)}$ is a block sub-matrix of $X$ obtained by dividing $X$ into $r$ by $r$ partitions. For instance, for $r = 2$,
\begin{equation}
X=
\left[
        \begin{array}{c;{2pt/2pt}c}
        X^{(1, 1)} & X^{(1, 2)} \\ \hdashline[2pt/2pt]
        X^{(2, 1)} & X^{(2, 2)}
    \end{array}
\right].
\end{equation}
An interpretation of $X'$ is that its $(k, l)$-th entry gives the amplitude and phase of the oscilatory component at frequency $\omega'=\left(\frac{2\pi}{n/r} k, \frac{2\pi}{m/r} l\right)$. Eq.~(\ref{eq:downsampling}) show this  $(k, l)$-th entry is the mean of $r^2$ different entries from $X$, and only one out of these $r^2$ different entries in $X$ correspond to the frequency $\omega = \left(\frac{2\pi}{n/r} k, \frac{2\pi}{m/r} l\right)$.
The other entries correspond to other frequencies, but, after downsampling, they will all be summed and it will not be possible to separate the contribution of each.

\section{Aliasing in CNNs}
\label{sec:aliasing-CNNs}

Convolutional layers are, in principle, capable of implementing anti-aliasing filter banks. Hence, even if anti-aliasing mechanisms are not directly encoded in the structure, CNNs could learn them and prevent aliasing from occurring in the intermediate signals.  One question is: ``do they actually learn such filter banks?'' In this section, we assess whether fully trained CNNs prevent aliasing from occurring for inputs taken from: i) the distribution the model has been trained on; and, ii) an adversarial distribution.

\subsection{Methodology: quantifying aliasing}
\label{sec:aliasing_methodology}

Here, we describe the methodology used to quantify the occurrence of aliasing in the intermediate layers of the CNN. Let $x$ and $x'$  be the signals before and after the downsampling operation and $X$ and $X'$ the corresponding DFT. To each value of $X'$ we attribute one of four mutually exclusive categories \{\textcolor{nopass}{no pass}, \textcolor{blue}{non-aliased}, \textcolor{aliased}{aliased}, \textcolor{aliasedtangled}{aliased-tangled}\}, described below. Fig.~\ref{fig:aliasing_in_CNNs} exemplify how the DFT entries of the CNN intermediate signals can be classified into these categories. We perform the analysis for all the layers in the CNN where the signal is downsampled: the strided convolutional layers are decomposed into normal convolutions followed by a downsampling operation; and, max-pooling layers, into a dense evaluation of max pooling (stride=1) followed by a downsampling operation. Then, we perform the analysis considering the signals immediately before and after the downsampling operation. We do that for all the channels.

Let $X^{(i, j)}$ be the block matrices that sum up to $X'$ in Eq.~(\ref{eq:downsampling_fft}) and fix a threshold value $T$. For instance, in the example below, we use $T=\max_{j, l}\|X'_{j, l}\|/10$. We say the $(p, q)$-th entry of the block $(i, j)$ \textit{contribute significantly to $X'_{p, q}$}  whenever $\|X^{(i, j)}_{p, q}\| >  T$. We classify the entry $(p, q)$ as \{\textcolor{nopass}{no pass}\} if no entry from the block matrices contribute significantly to $X'_{p, q}$. Hence, these frequency components have a relatively small magnitude and removing them would have little to no effect on the signal $x'$.
\begin{figure}[t]
    \vspace{-7pt}
    \centering
    \begin{tabular}{cc}
        \hspace{-20pt}
        \begin{subfigure}{0.26\textwidth}
            \includegraphics[width=\linewidth]{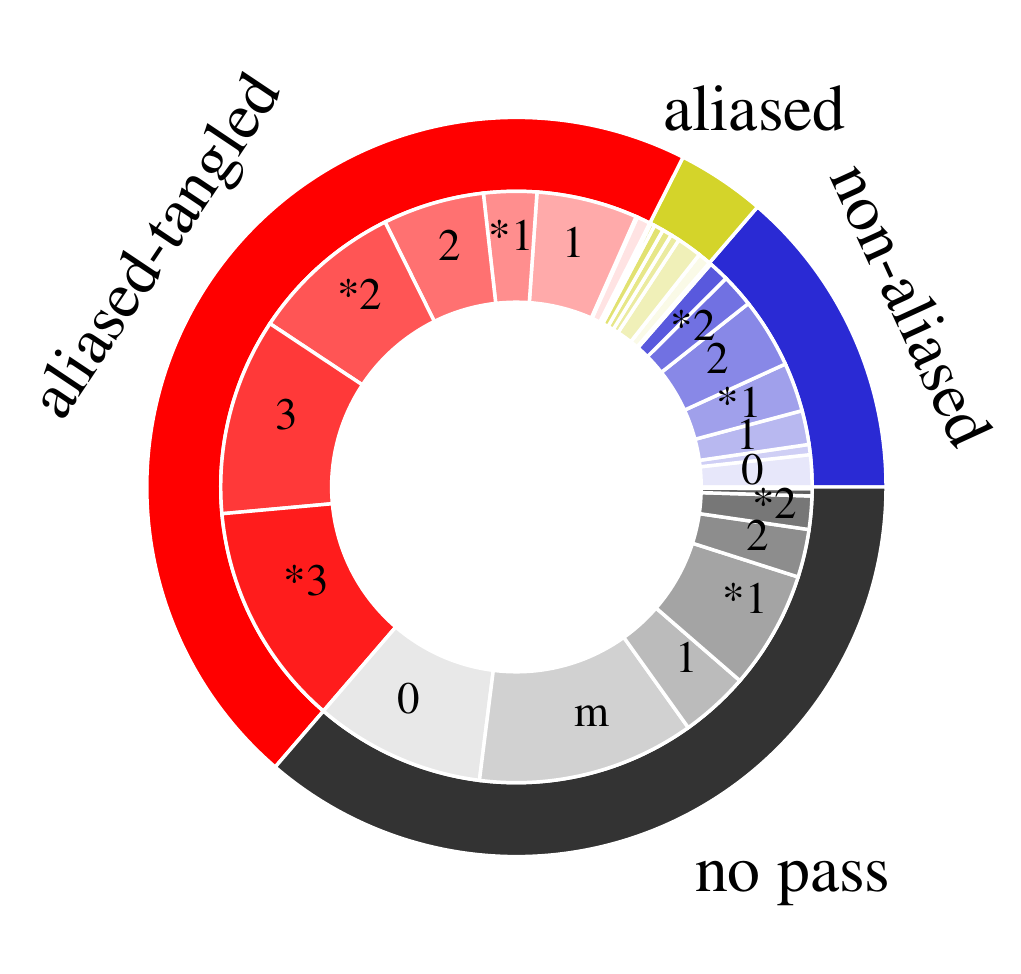}
            \caption{ImageNet.}
            \smallskip
         \end{subfigure}
        \begin{subfigure}{0.26\textwidth}
            \includegraphics[width=\linewidth]{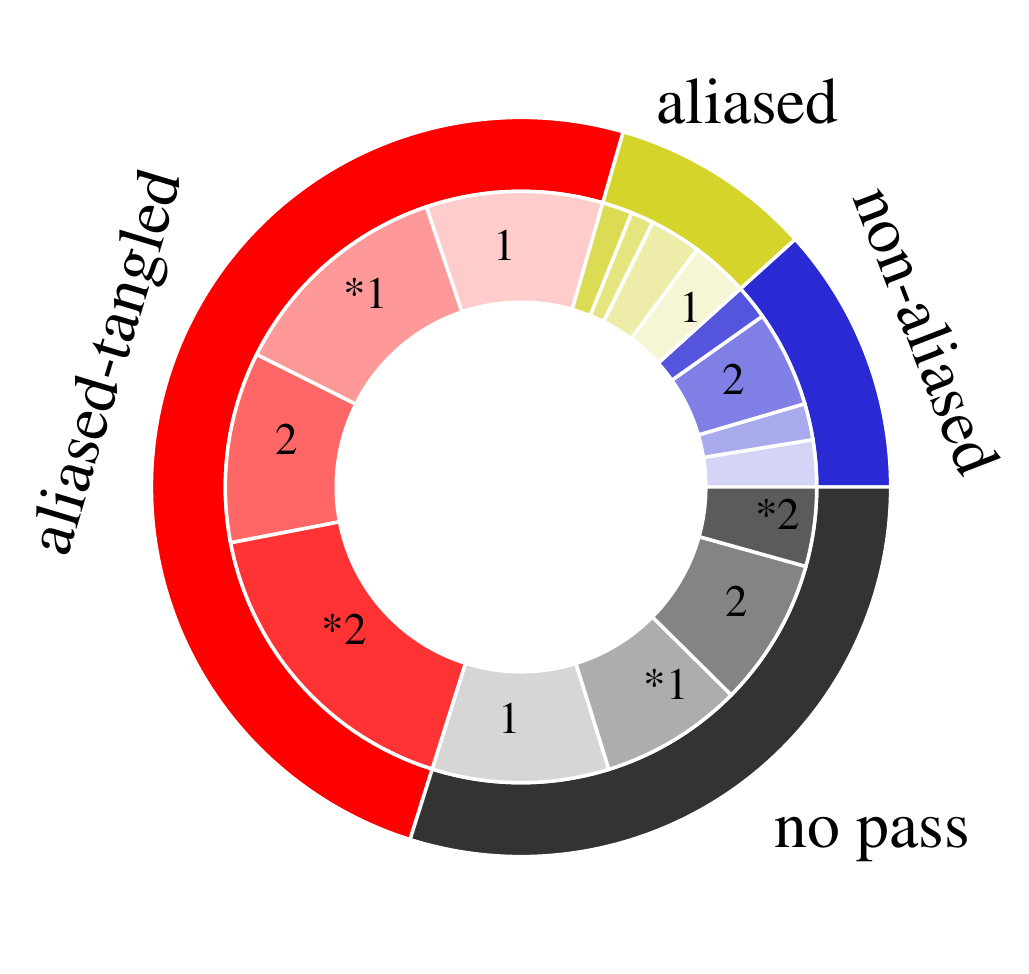}
            \caption{Oscilation classif.}
            \smallskip
        \end{subfigure}
        \end{tabular}
    \vspace{-5pt}
    \caption{\textbf{Fraction of samples suffering aliasing}. Pie chart indicating the fraction of intermediate signals (immediately after the downsampling operation) in each of the categories described in Section~\ref{sec:aliasing_methodology}. The inner chart distinguish between different downsampling operations (with the opacity specifying the layer: more opaque corresponding to signals in the deeper layers of the network) that take place over the neural network. The outer chart show the average fraction over all. Both for: (a) ResNet34 on ImageNet; and (b), for Resnet20 on the task described in Section~\ref{sec:classifying-oscilations}. }
   \label{fig:fraction-of-aliasing}
\end{figure}

Now, assume that at least one block matrix contributes significantly to $X'_{p, q}$. If the block matrix contributing is $(i, j) = (1, 1)$, and only this one, we say that this is a \{\textcolor{blue}{non-aliased}\} entry, since the signal distortion due to aliasing is small (or zero). If a single block $(i, j) \not= (1, 1)$ contribute significantly to $X'_{p, q}$, we say that this entry is \{\textcolor{aliased}{aliased}\}.  If more than one block contributes significantly to $X'_{p, q}$ we say the entry is \{\textcolor{aliasedtangled}{aliased-tangled}\}. We distinguish between these two types of aliasing because, while in the first case a high-frequency component might appear as a low-frequency component after downsampling, the reconstruction of the original component could be made easy by the knowledge that this has happened and of which block have contributed significantly to this point.  In the second case, multiple frequency components are summed together, making the reconstruction harder due to the need of distinguishing between multiple additive components.

\begin{figure}[t]
    \vspace{-8pt}
    \centering
    \includegraphics[width=0.93\linewidth]{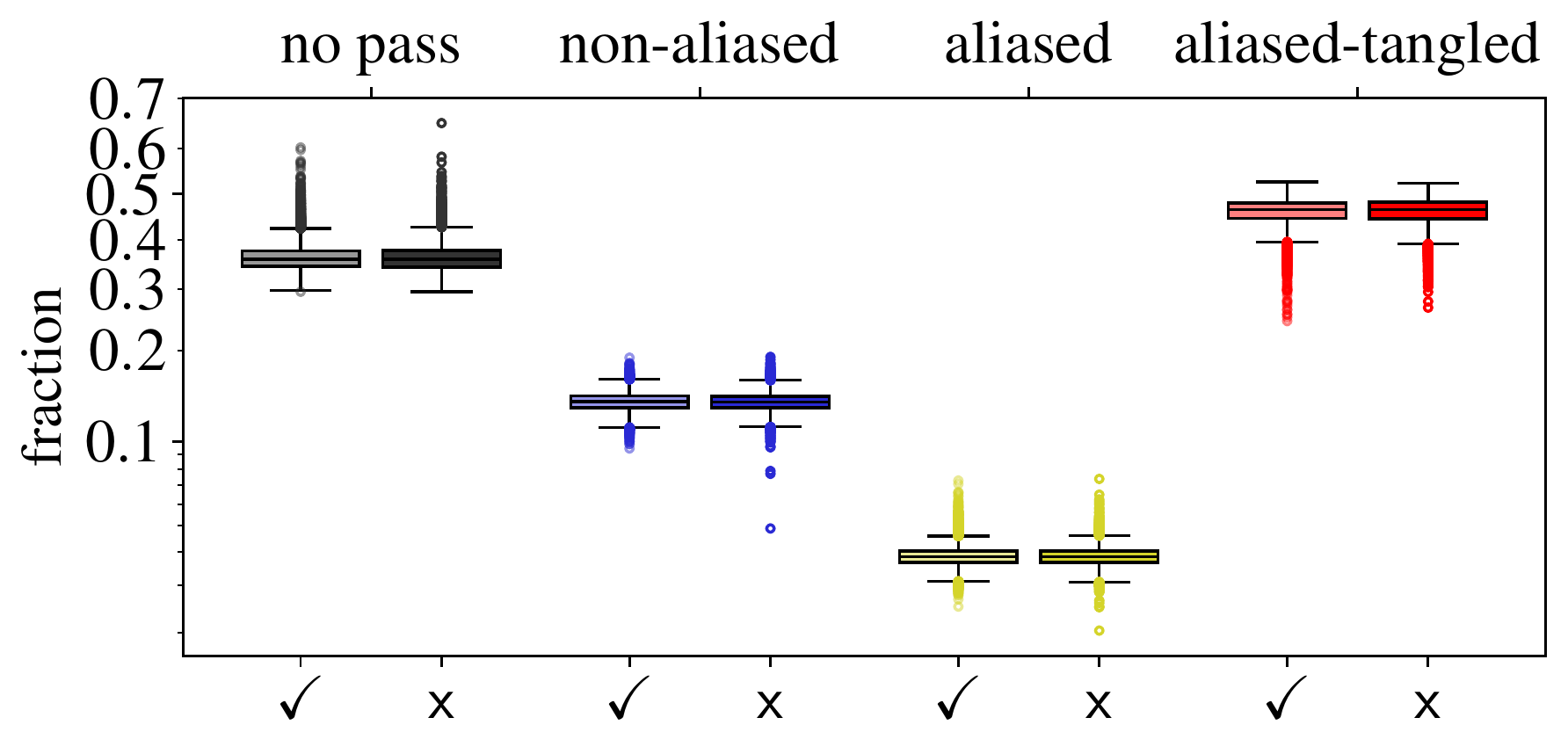}
    \vspace{-7pt}
    \caption{\textbf{Correct ({\sffamily x}) \textit{vs} incorrect ($\checkmark$)} classified examples in ImageNet test set. Each test sample is a data point. The fraction of pixels in each category is displayed on the $y$-axis, and on the $x$-axis we have the categories described in Section~\ref{sec:classifying-oscilations}, further subdivided into correctly and incorrectly classified examples.}
   \label{fig:correct-vs-incorrect}
   \vspace{-10pt}
\end{figure}

\subsection{Results}
\label{sec:aliasing_results}
For ImageNet~\cite{krizhevsky_imagenet_2012} experiments we use the residual neural network~\cite{he_deep_2016} with 34 layers (ResNet34) available in the \textit{torchvision} library~\cite{paszke_pytorch_2019}. The model was trained on ImageNet standard training set and achieves---on the test set---73.3\% accuracy and 91.4\% \hbox{top-5} accuracy (i.e., the percentage of cases where one of the 5 highest probability model predictions matches the ground truth).

\noindent{$\bullet$ \textbf{ImageNet classification:}}
We observe the intermediate signals of the ResNet34 when evaluated on samples from ImageNet test set. The ResNet34 downscales the signal in strided convolutional and maxpooling layers in a total of 7 different intermediate points, we denote the intermediate signals immediately after these downsample points by $\{0, \text{m}, 1, *1, 2, *2, 3, *3\}$ (here we use `*' to denote the signal in the skip connection). Fig.~\ref{fig:aliasing_in_CNNs} shows the DFT of two intermediate signals (\{0\} and \{1\}), with the frequency components classified according to Section~\ref{sec:aliasing_methodology}.

Fig.~\ref{fig:fraction-of-aliasing}(a) displays the fraction of points in each category averaged over all channels and over all examples in the ImageNet test set. The outer chart shows the---equally weighted---average over the 7 intermediate signals under consideration and the inner chart shows the individual contributions of each intermediate signal to the average. Aliasing occurs in, roughly, half of the frequency components in intermediate layers, with a predominance of what we call \textcolor{aliasedtangled}{aliased-tangled}, which yields a harder reconstruction problem. The fraction of aliased  components seems to be more densely concentrated in the final layers of the network. The initial layers contain more frequency components with relatively small magnitude (\textcolor{nopass}{no pass}).

 \begin{figure}[t]
    \vspace{-4pt}
    \centering
    \includegraphics[width=0.93\linewidth]{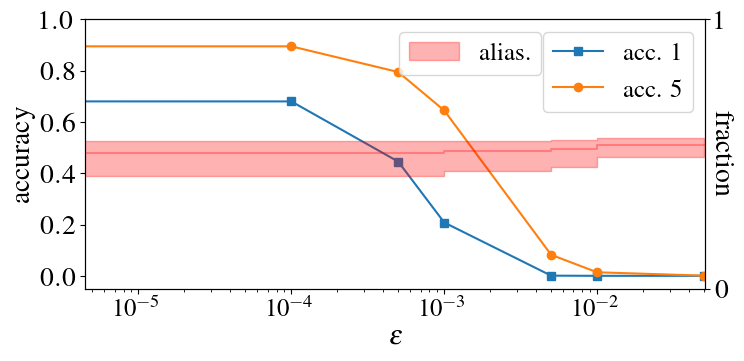}
    \vspace{-7pt}
    \caption{\textbf{Adversarial samples}. We display the (top-1) accuracy and the top-5 accuracy as a function of $\epsilon$ used in generating the adversarial examples. Also as a function of $\epsilon$, we display the number of intermediate components that suffer aliasing (\textcolor{aliased}{aliased} + \textcolor{aliasedtangled}{aliased-tangled}): the  full line is the median and the shaded region is between the 1\% and 99\% percentiles. We use 100 steps of projected gradient descent with maximum step size ($2.5\epsilon /100$) to generate the adversarial disturbance as proposed in~\cite{madry_towards_2018}.}
   \label{fig:adversarial-example}
    \vspace{-6pt}
\end{figure}

\noindent{$\bullet$ \textbf{Aliasing when classifying oscillations:}}
Fig.~\ref{fig:fraction-of-aliasing}(b) shows the same experiment for a residual neural network with 20 layers (ResNet20) used for the task described in Section~\ref{sec:classifying-oscilations}. The model has only 4 intermediate downsampling points $\{1, *1, 2, *2\}$ and achieves an accuracy of $99.2\%$ (for noise intensity $A=1$). The task requires the neural network to be able to distinguish between the input frequencies, and while the model succeeds at the task, it does not do so by explicitly implementing anti-aliasing mechanisms, since more than half of the frequency components of the intermediate signals suffer aliasing.

\noindent{$\bullet$ \textbf{Aliasing in correct and incorrect classified instances:}}
In Fig.~\ref{fig:correct-vs-incorrect}, we compare the fraction of aliased components for correct \textit{vs} incorrect classified examples in ImageNet. The obtained distribution is almost identical for correctly and incorrectly classified examples. This, in turn, suggests that, during inference, the occurrence of aliasing in a CNN has little to do with its capability of classifying images from the test distribution. 

\noindent{$\bullet$ \textbf{Aliasing in adversarial examples:}}
We also analyse the classification of images coming from an adversarial distribution. That is, instead of an input $x$ from the test set, the neural network will be fed an input $x + v$, being the vector $v$ chosen adversarially, in the region $\|v\|_{\infty} \le \epsilon$, to yield the worst possible model prediction (i.e., for which the loss function achieves its highest value). Fig.~\ref{fig:adversarial-example} contain the accuracy as a function of $\epsilon$ together with the fraction of frequency components that actually suffer aliasing in the intermediate signals. While there is a small increase in the fraction of aliased components, the small magnitude of it does not seem to point to aliasing as the main weakness explored by the adversarial attack.

\section{Discussion}
\label{sec:discussion}

Nyquist's sampling theorem~\cite{nyquist_certain_1928} gives a \textit{sufficient} criterion for guaranteeing the possibility of reconstructing signals. In light of this theorem, anti-aliasing filters are a practical solution to downsampling without losing information. Nonetheless, later developments in signal processing have shown that it is possible to reconstruct signals sampled below the Nyquist rate given that the signal is sparse in some domain~\cite{candes_decoding_2005, donoho_compressed_2006}. In the case of CNNs, despite aliasing taking place in intermediate signals (see Section~\ref{sec:aliasing-CNNs}) the possibility of reconstruction is simplified by the channel redundancy, which \textit{might} play an important role in distinguishing between frequency components, cf. Section~\ref{sec:classifying-oscilations}.

\appendix


\begin{thebibliography}{10}

\bibitem{le_cun_handwritten_1990}
Yann Le~Cun, Ofer Matan, Bernhard Boser, John~S Denker, Don Henderson,
  Richard~E Howard, Wayne Hubbard, LD~Jacket, and Henry~S Baird,
\newblock ``Handwritten zip code recognition with multilayer networks,''
\newblock in {\em Proceedings of the 10th {{International Conference}} on
  {{Pattern Recognition}} ({{ICPR}})}. {IEEE}, 1990, vol.~2, pp. 35--40.

\bibitem{scherer_evaluation_2010}
Dominik Scherer, Andreas Muller, and Sven Behnke,
\newblock ``Evaluation of {{Pooling Operations}} in {{Convolutional
  Architectures}} for {{Object Recognition}},''
\newblock in {\em Artificial {{Neural Networks}} \textendash{} {{ICANN}}}, vol.
  6354, pp. 92--101. {Springer}, {Berlin, Heidelberg}, 2010.

\bibitem{he_deep_2016}
Kaiming He, Xiangyu Zhang, Shaoqing Ren, and Jian Sun,
\newblock ``Deep {{Residual Learning}} for {{Image Recognition}},''
\newblock in {\em Proceedings of the {{IEEE Conference}} on {{Computer Vision}}
  and {{Pattern Recognition}} ({{CVPR}})}, 2016, pp. 770--778.

\bibitem{zhang_making_2019}
Richard Zhang,
\newblock ``Making {{Convolutional Networks Shift}}-{{Invariant Again}},''
\newblock in {\em Proceedings of the 36th {{International Conference}} on
  {{Machine Learning}} ({{ICML}})}, June 2019.

\bibitem{azulay_why_2019}
Aharon Azulay and Yair Weiss,
\newblock ``Why do deep convolutional networks generalize so poorly to small
  image transformations?,''
\newblock {\em Journal of Machine Learning Research}, vol. 20, pp. 1--25, 2019.

\bibitem{tanaka_wavecyclegan2_2019}
Kou Tanaka, Hirokazu Kameoka, Takuhiro Kaneko, and Nobukatsu Hojo,
\newblock ``{{WaveCycleGAN2}}: {{Time}}-domain {{Neural Post}}-filter for
  {{Speech Waveform Generation}},''
\newblock {\em arXiv:1904.02892}, Apr. 2019.

\bibitem{henaff_geodesics_2016}
Olivier~J. H{\'e}naff and Eero~P. Simoncelli,
\newblock ``Geodesics of learned representations,''
\newblock in {\em Proceedings of the {{International Conference}} for
  {{Learning Representations}} ({{ICLR}})}, Feb. 2016.

\bibitem{vasconcelos_effective_2020}
Cristina Vasconcelos, Hugo Larochelle, Vincent Dumoulin, Nicolas~Le Roux, and
  Ross Goroshin,
\newblock ``An {{Effective Anti}}-{{Aliasing Approach}} for {{Residual
  Networks}},''
\newblock {\em arXiv:2011.10675}, Nov. 2020.

\bibitem{zou_delving_2020}
Xueyan Zou,
\newblock ``Delving {{Deeper}} into {{Anti}}-aliasing in {{ConvNets}},''
\newblock in {\em Proceedings of the 31st {{British Machine Vision Virtual
  Conference}} ({{BMVC}})}, 2020.

\bibitem{kingma_adam:_2014}
Diederik~P. Kingma and Jimmy Ba,
\newblock ``Adam: {{A Method}} for {{Stochastic Optimization}},''
\newblock in {\em Proceedings of the 3rd {{International Conference}} for
  {{Learning Representations}} ({{ICLR}})}, Dec. 2014.

\bibitem{gong_impact_2018}
Yuan Gong and Christian Poellabauer,
\newblock ``Impact of {{Aliasing}} on {{Deep CNN}}-{{Based End}}-to-{{End
  Acoustic Models}},''
\newblock in {\em Interspeech}, 2018, pp. 2698--2702.

\bibitem{krizhevsky_imagenet_2012}
Alex Krizhevsky, Ilya Sutskever, and Geoffrey~E. Hinton,
\newblock ``Imagenet classification with deep convolutional neural networks,''
\newblock in {\em Advances in {{Neural Information Processing Systems}}}, 2012,
  pp. 1097--1105.

\bibitem{paszke_pytorch_2019}
Adam Paszke, Sam Gross, Francisco Massa, Adam Lerer, James Bradbury, Gregory
  Chanan, Trevor Killeen, Zeming Lin, Natalia Gimelshein, Luca Antiga, Alban
  Desmaison, Andreas Kopf, Edward Yang, Zachary DeVito, Martin Raison, Alykhan
  Tejani, Sasank Chilamkurthy, Benoit Steiner, Lu~Fang, Junjie Bai, and Soumith
  Chintala,
\newblock ``{{PyTorch}}: {{An}} imperative style, high-performance deep
  learning library,''
\newblock in {\em Advances in Neural Information Processing Systems 32}, pp.
  8024--8035. 2019.

\bibitem{madry_towards_2018}
Aleksander Madry, Aleksandar Makelov, Ludwig Schmidt, Dimitris Tsipras, and
  Adrian Vladu,
\newblock ``Towards {{Deep Learning Models Resistant}} to {{Adversarial
  Attacks}},''
\newblock {\em Proceedings of the International Conference for Learning
  Representations (ICLR)}, 2018.

\bibitem{nyquist_certain_1928}
Harry Nyquist,
\newblock ``Certain topics in telegraph transmission theory,''
\newblock {\em Transactions of the American Institute of Electrical Engineers},
  vol. 47, no. 2, pp. 617--644, 1928.

\bibitem{candes_decoding_2005}
Emmanuel~J Candes and Terence Tao,
\newblock ``Decoding by linear programming,''
\newblock {\em IEEE transactions on information theory}, vol. 51, no. 12, pp.
  4203--4215, 2005.

\bibitem{donoho_compressed_2006}
David~L. Donoho,
\newblock ``Compressed sensing,''
\newblock {\em IEEE Transactions on Information Theory}, vol. 52, no. 4, pp.
  1289--1306, 2006.

\end{thebibliography}
\end{document}